# *Cellulyzer*

## Automated analysis and interactive visualization/simulation of select cellular processes

Aliakbar Jafarpour and Holger Lorenz

*Zentrum für molekulare Biologie der Universität Heidelberg (ZMBH), INF 282, 69120 Heidelberg, Germany*

*jafarpour.a.j@ieee.org*

*h.lorenz@zmbh.uni-heidelberg.de*

**Abstract:** Here we report on a set of programs developed at the ZMBH Bio-Imaging Facility for tracking real-life images of cellular processes. These programs perform 1) automated tracking; 2) quantitative and comparative track analyses of different images in different groups; 3) different interactive visualization schemes; and 4) interactive realistic simulation of different cellular processes for validation and optimal problem-specific adjustment of image acquisition parameters (tradeoff between speed, resolution, and quality with feedback from the very final results). The collection of programs is primarily developed for the common bio-image analysis software *ImageJ* (as a single Java Plugin). Some programs are also available in other languages (C++ and Javascript) and may be run simply with a web-browser; even on a low-end Tablet or Smartphone. The programs are available at

https://github.com/nurlicht/CellulyzerDemo

## 1. Introduction

Applying a tracking program to a time-series image of moving particles generates a table of tracks, as shown in Figure 1. Associated with each track in this table are some rows including the parameters of individual track points. Achieving such a table is the perceived (initial) goal of a biologist having imaged a cellular process, and it is possible with many *general-purpose* tracking programs.

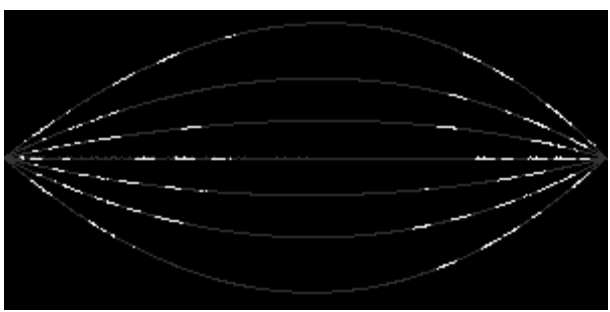

| Track | Point | X | Y | Frame | . |
|---|---|---|---|---|---|
| 1 | 1 | 20 | 80 | 1 | |
| 1 | 2 | 22 | 84 | 2 | |
| 1 | 3 | 25 | 90 | 3 | |
| 2 | 1 | 26 | 70 | 2 | |
| 2 | 2 | 29 | 66 | 3 | |
| . | . | . | . | . | |
| 237 | 4 | 240 | 60 | 130 | . |

**Figure 1:** (Left) Simulated snapshot of a cell-biological process (mitosis), and (Right) the corresponding output of tracking; i.e., the list of tracks and the properties of their constituent points

*Requirements Engineering* of common tracking problems in cell biology shows that achieving a table of tracks is not the end, and also not the beginning. The following requirements are also important aspects of tracking problems for the higher-level problem of fast, clear, (maximally-) reproducible, and (maximally-) objective biological conclusions:
- Validation and intuitive interpretation of tracking results,

- Problem-specific required analyses before (and coupled with) tracking,
- The possibility of enforcing prior qualitative biological knowledge,
- Problem-specific analyses after tracking, and
- Systematic merge/comparison/representation of the results from different experiments.

Such considerations are different from those in other tracking contexts (robotics, surveillance, autonomous cars, advanced driver-assistance systems …) and the mentality behind the programs developed therein.

The programs offered here have emerged in the course of two years of communication and collaboration with cell biologists (especially in handling videos of microtubule plus-end proteins and centrosomes),

- To understand the ultimate goals and the requirements of tracking,
- To identify the gaps between existing tracking programs and practical problems,
- To formulate the problems mathematically and to develop pertinent solutions,
- To design the pertinent software architectures and algorithms,
- To implement and to test the required programs,
- To separate mathematical/programming technicalities/terminologies from the end-user, yet
- To give the end-user flexible, intuitive, and interactive tools to visualize, to validate, and to select the (sought) results, and
- To enable the user to simulate *realistic* test images interactively
  - To choose the right microscope parameters before conducting an experiment (for easy validation and efficient use of biological/optical resources and man-hours), and
  - To validate analyses using test images with known user-defined parameters (*Black-Box Testing*).

This report has been structured as follows: *Section 2* exemplifies specific goals of tracking in select cell-biological problems. An overview of the programs and programming aspects are given in *Sections 3* and *4*, respectively. Some practical aspects are discussed in *Section 5*, and brief conclusions are made in *Section 6*. *Appendix I* explains dialog-boxes seen by the end-user, and *Appendix II* addresses some practical issues in track analyses. Finally, mathematical and architectural aspects of the programs are detailed in *Appendices III* and *IV*, respectively.

## 2. Specific goals of tracking in select cell-biological problems
### 2.1. Comparative study of asymmetry

Figure 2 (top-left) shows a snapshot of particles in a simulated cell-reproduction process (microtubule-associated end-binding proteins of a mitotic spindle). Parameters such as speed and duration (lifetime) of such particles are commonly studied quantitatively (Chinen, 2015). A less common quantification scheme concerns the angular orientations of microtubule tails, which is sometimes reported for a single experiment; i.e., without merging the results of different experiments [Figure 1.d in (Kleele, 2014) and (Miura, 2016)].

Our goal in tracking is to reduce the information content of orientations to a few conclusive indices, so that different time-series images can be easily, reliably, and quantitatively compared (irrespective of size, position, or absolute orientations and without registration). This goal is achieved using the two *directionality* and *asymmetry* indices, as defined in Figure 2.

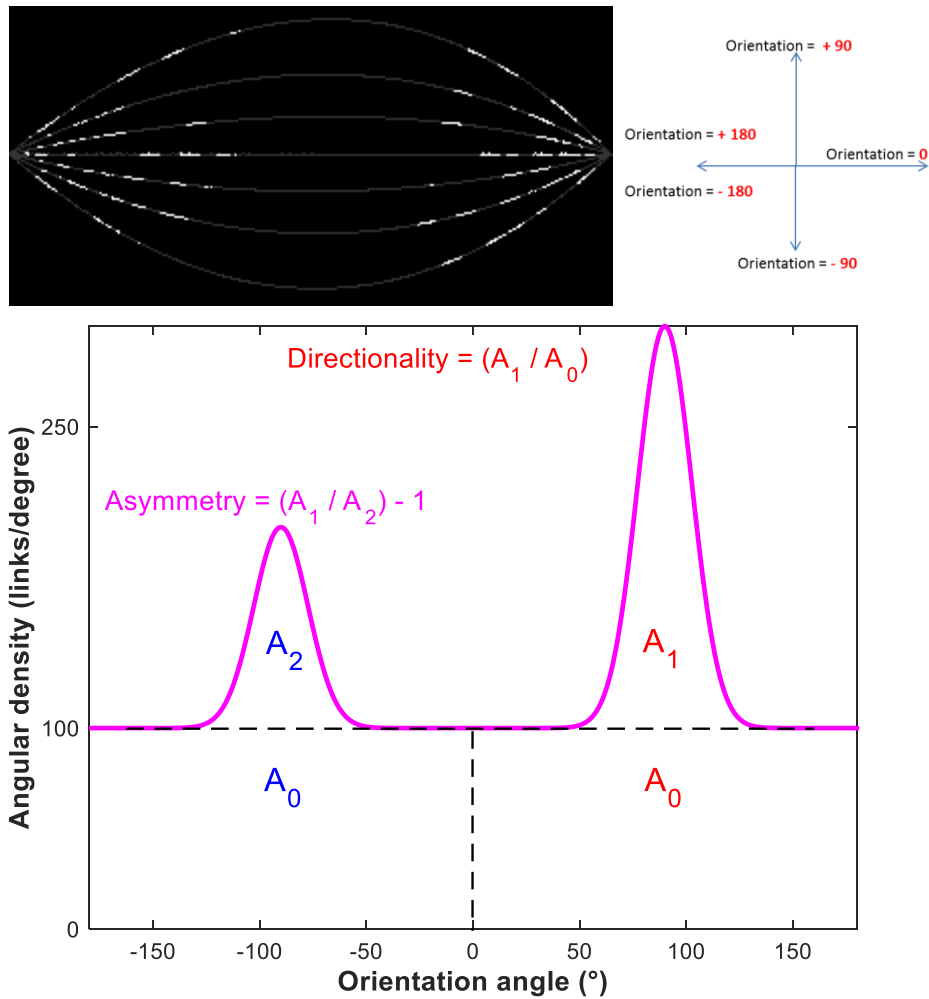

**Figure 2:** (Top-left) A snapshot of particles in a simulated cell-reproduction process (microtubule-associated end-binding proteins of a mitotic spindle). The particles move along different orientations, which can be quantified according to the common convention (top-right). The combined orientations of all links along all tracks give rise to a histogram (bottom). In such an orientation histogram, the horizontal axis represents angles (in degrees), and the vertical axis represents the number of detected links (two consecutive spots along a given track) per unit of angle. The two *directionality* and *asymmetry* indices summarize and quantify the orientation information in a comparable and concise way. As such, they make it possible to compare different such processes automatically without registration.

## 2.2. Pair-tracking

Figure 3 shows consecutive frames of a simulated centrosome-pair time-series. The goal of tracking is to answer questions such as "How does the histogram of all intra-pair distances look like? How can it be quantified with intuitive and comparable indices? How often does switching (between *Small* and *Large* distances) occur? How are such properties affected in mutated cells compared to normal cells? …"

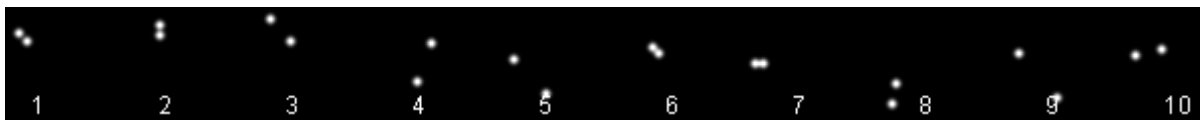

**Figure 3:** Consecutive frames of a simulated centrosome-pair time-series with dynamic stochastic intra-pair distances.

## 2.3. Diffusion-related analyses

Figure 4 (top) shows in the background the simulated diffusion trajectories of three particles. The goal of tracking is to identify different *diffusion regimes* and to visualize (different subsets of) the trajectories based on the associated regimes.

Superimposed on the image in the foreground (with yellow spots) are parts of these trajectories that correspond to the chosen diffusion regime. The number and the ranges of different diffusion regimes, the critical length of sub-trajectories (Duchesne, 2012), and some other parameters can be adjusted interactively.

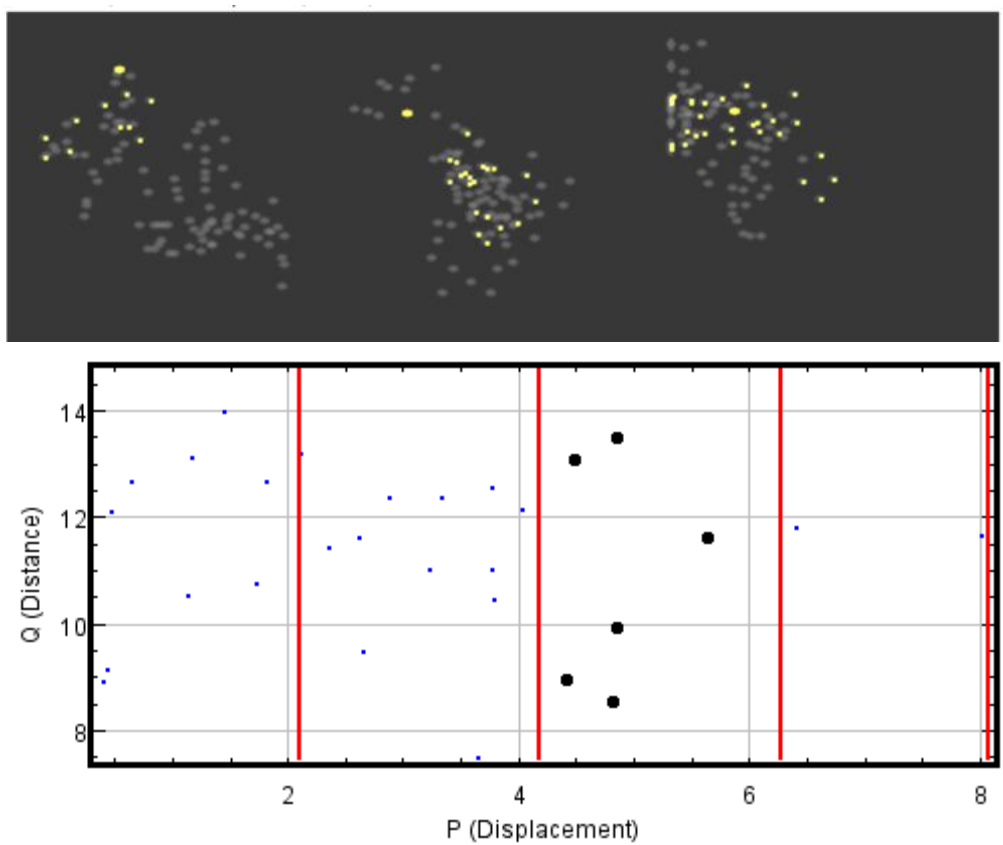

**Figure 4:** Simulated diffusion trajectories of three particles (top) and the associated "Displacement-Distance" plot of sub-trajectories (bottom). Different intervals in the bottom plot are (user-defined) *regimes of diffusion*. It is possible to select, and to analyze only sub-trajectories that belong to the same regime of diffusion, as shown with yellow spots (top) and black spots (bottom).

## 2.4. Single-cell statistics

Figure 5 shows simulated particles emanating from two adjacent sources, before the start of the cell-reproduction process (microtubule-associated end-binding proteins in *Interphase*). The goal of tracking is to first associate the tracks to the corresponding sources, and then to calculate different track properties (speed, number of tracks …) at single-cell (single-source) level.

Here, many particles are close to the opposite source and/or in close proximity of particles emanating from this source. Consistent, easily-adjustable, and easily-verifiable measures are required to associate such particles (and their associated tracks) to the corresponding sources.

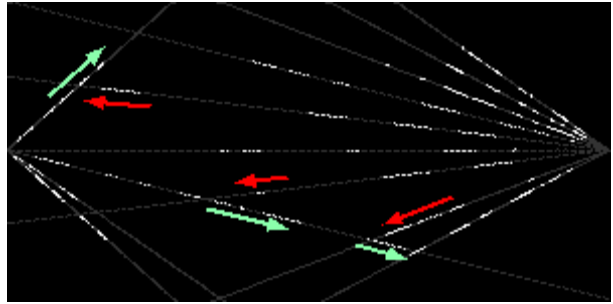

**Figure 5:** Simulated particles emanating from two sources, before the start of the cell-reproduction process (microtubule-associated end-binding proteins in *Interphase*). Particles next to red/green arrows emanate from the source on the right/left.

## 3. Overview of the programs

Figure 6 shows the block diagram of the collective program; featuring a typical analysis pipeline (feedforward path) and also the implicit (feedback path for) adjustment of algorithm and image parameters. The core of the code can be tested or further developed even without measured data (using *realistic* simulated images). The image acquisition parameters can also be determined systematically with problem-specific feedback from the very final results (rather than using "generally-optimal" or intuitively-selected values and wasting biological/optical resources in try-and-error runs).

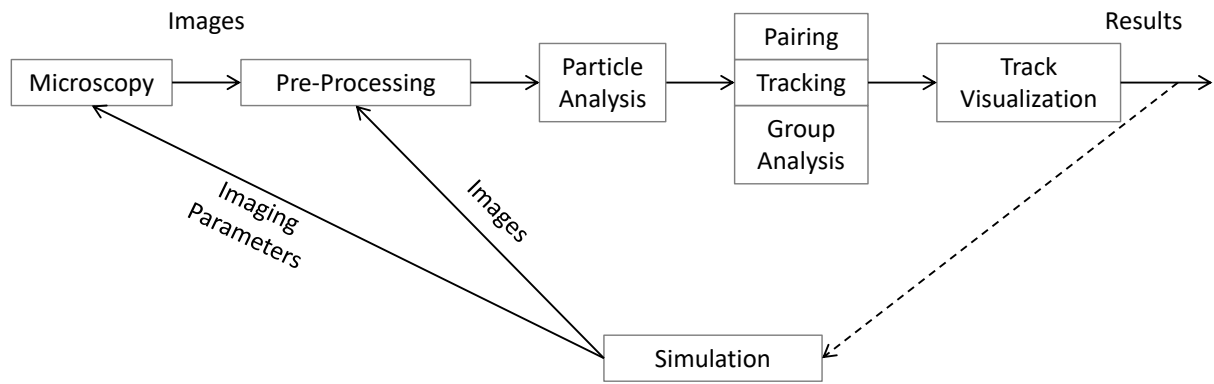

**Figure 6:** The block diagram of the collective program featuring a typical analysis pipeline (feedforward path) and also the implicit (feedback path for) adjustment of algorithm and image parameters.

The abstract operations of Figure 6 are seen by the end-user, as depicted in Figure 7. Dialog-boxes related to different operations and the meanings of different options have been detailed in *Appendix I*.

Tracking is first facilitated by appropriate *pre-processing* to enhance the visibility of the sought particles in comparison with unwanted objects. Quantitative particle analysis helps the user to find optimal spot-detection parameters interactively. These parameters will be needed later for optimal automated tracking. If needed (as in the case of centrosome pairs), identified spots in each frame are first categorized into pairs, before tracking. Tracking can be done either by Fiji's *TrackMate* functionality or with the same algorithm used for pairing (this time linking nearby spots *across* frames, rather than nearby spots *in the same* frame).

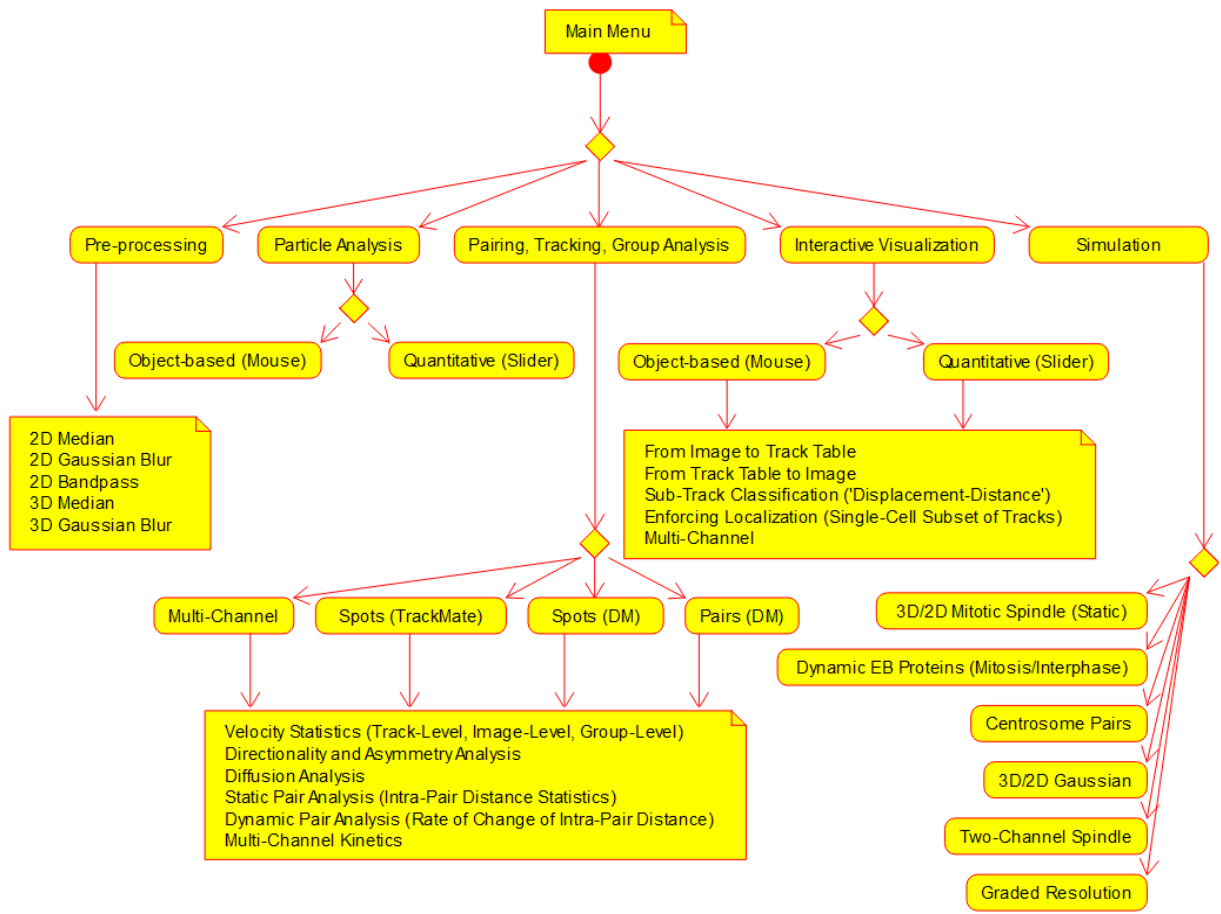

**Figure 7:** This *Activity Diagram* correlates the abstract operations of the blocks in Figure 6 to the multiple-choice dialog-boxes observed by running the program. Rhombus-shaped elements denote a choice of a unique operation (out of two or more). Rectangular boxes (with folded upper-right corner) denote multiple such operations (that might be performed together). The dialog-boxes offering these options all feature a "Help" button for a quick introduction of the operations.

In addition to common track analyses (velocity, density …), three specific categories of Pair Analysis, Asymmetry (Directionality) Analysis, and Diffusion Analysis have also been implemented. Pair Analysis includes statistics of intra-pair distances (static) and also the rate of change of these distances (dynamic). Asymmetry Analysis is based on the histogram of track orientations, from which two figures of merit (*directionality* and *asymmetry*) are calculated.

In both cases of Pair Analysis and Orientation Analysis, each time-series image is reduced to two figures of merit. These two numbers are calculated for different images in different groups and demonstrated (with appropriate color-coding based on group index) in a single scatter plot.

Diffusion Analysis includes calculation of Mean Square Displacement (MSD) vs. time-lag. Further investigation of diffusion is possible using interactive "Displacement-Distance" visualization, as shown in Sec. 2.3.

Different modes of track visualization help the user to visualize, validate, and select tracks (and track subsets) interactively. Interactive simulations make it possible to synthesize realistic images for validation and test of tracking results, and to find the optimal image acquisition parameters before doing experiments.

## 4. Programming aspects
### 4.1. Previous works
Two special-purpose tracking programs for such problems are reported in (Sironi, 2011) and (Applegate, 2011); both written with the commercial program Matlab (*The Mathworks, Inc.*). The common program ImageJ (*The National Institute of Health*) is also employed for basic processing (filtering, Region of Interest selection …).

A more recent contribution (Soundararajan, 2014) uses the program *TrackMate*, which is part of the Fiji distribution of ImageJ (Tinevez, 2016 online). The results are then exported and further processed with Matlab.

*4.2. Current contribution*

The programs have been primarily developed (as a single Java Plugin) for the bio-image analysis software *ImageJ* . Individual programs have been developed by implementing the *PlugIn*, *PlugInFilter*, *ExtendedPlugInFilter*, and *DialogListener* interfaces offered by ImageJ.

The starting point for most interactive programs has been the *Cross_Fader.java* program, developed by Michael Schmid (Schmid, 2014). In further development of such programs, specific patterns and similarities have been observed, and the architecture has been modified accordingly as detailed in *Appendix IV*. Developing a few cases of multi-threaded loops (in the class *Tracking*) has been based on the sample codes by Stephan Preibisch and Albert Cardona (Cardona, 2012).

Some programs are also available as C++ or Javascript implementations. Javascript programs (and C++ programs ported to Chrome *Native Client*) can be run just with a web-browser on a low-end Tablet or Smartphone.

|  | **Functionality** | **Interactivity** | **Required API** |
|---|---|---|---|
| **Java** | Full | Full (mouse, sliders …) | ImageJ (Fiji) |
| **IJ Macro** | Tracking, pairing, and track analyses | Mouse, Dialog boxes | ImageJ (Fiji) |
| **C++** | Tracking | Sliders | OpenCV |
| **Javascript** | Time-Series Simulation | Slider-like | (Built-in) Canvas |

**Table 1:** Implementation of the programs in different programming languages and the corresponding functionality, interactivity, and required Application Programming Interfaces (APIs).

## 5. Discussions

*5.1. When does automated tracking fail (even theoretically)?*

The critical parameter in answering this question is the Displacement-to-Distance Ratio, DDR. Here Displacement refers to the movement of a particle *across* two frames, whereas Distance refers to the separation between particles *in* a single frame.

If the frame-rate is high enough and/or the particles are sparse enough, then automated tracking is preferred. This situation corresponds to small values of DDR; DDR < 0.5.

As the DDR increases (Figure 8), automated tracking becomes more problematic. Note that it is a matter of lack of information (poor temporal resolution) in data, and it cannot be simply compensated with "better programs". Using prediction-correction (Kalman-filtering-type) schemes (Peterfreund, 2002) or the similarity of particles (not common in tracking bio-images) may theoretically retrieve some of the lacking information, but at the cost of other ambiguities and more difficult validation.

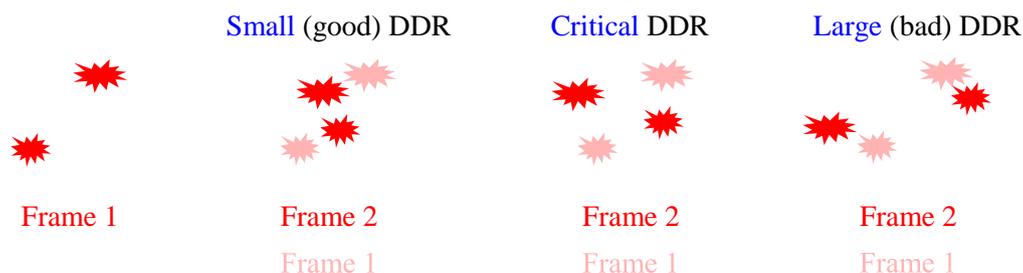

**Figure 8:** The first frame of a simulated time-series including two particles (left), and three potential scenarios for the next frame (right). A Particle will be trackable if its displacement (across two frames) is smaller than its distance to its neighbors (in a given frame). This property is quantified as Displacement-to-Distance Ratio or DDR; ideally DDR < 0.5.

For the common case of tracking dense end-binding proteins (Figure 1), we have found high frame rates (~7Hz; i.e., ~10 times higher than typically-employed values) crucial for reliable automated tracking (low DDR). This increased frame rate is associated with reduced spatial resolution, reduced exposure time, and reduced laser power (to minimize bleaching). Despite the reduced resolution and signal-to-noise ratio in individual images, spot-detection algorithms are robust enough to reliably detect particles. The reduced DDR, however, has been crucial for linking particles across frames reliably.

Even with low DDR, manual tracking or pre-processing images before tracking is required to handle non-trivial split-merge events such as cell-reproduction (mitosis). Explicit problem-specific biological meaning and analysis requirement are necessary here to deal with the problem, and "better programs" cannot handle these events on their own. Manual tracking is recommended here (at least if not exhaustive).

*5.2. Simulated vs. measured asymmetry*

Using different values of asymmetry in simulated time-series images of cell-reproduction (Figure 2) is expected to produce corresponding distinguished values in the measured asymmetry. Figure 9 shows such a correspondence or calibration curve.

The *measured* asymmetry is defined (in terms of the number of detectable links) as in Figure 2 and can assume arbitrarily large values. The *simulated* asymmetry (in terms of the number of particles) is defined with a maximum theoretical value of 0.5 (for compatibility with other types of asymmetry). The theoretical calibration curve correlating these two quantities is a hyperbolic segment, and resembles the observed curve in Figure 9 (semi-quantitatively).

Note that the simulated asymmetry is evaluated based on the slope of the underlying track at the origin (Microtubule Organizing Center, MTOC). As such, for all particles moving along the same underlying track, the simulated orientation is the same, whereas the measured orientation depends on the location along the track.

The underlying processes are stochastic, and simulated videos with the same expected value of simulated asymmetry are only in long-term and only in the statistical sense comparable. Nevertheless for the simulated 400-frame videos, the observed distribution of the measured asymmetries is relatively small. Aside from such expected statistical distributions, this diagram confirms that the figure of merit used for measured asymmetry is *quantitatively* meaningful and *in a statistical sense* "reproducible".

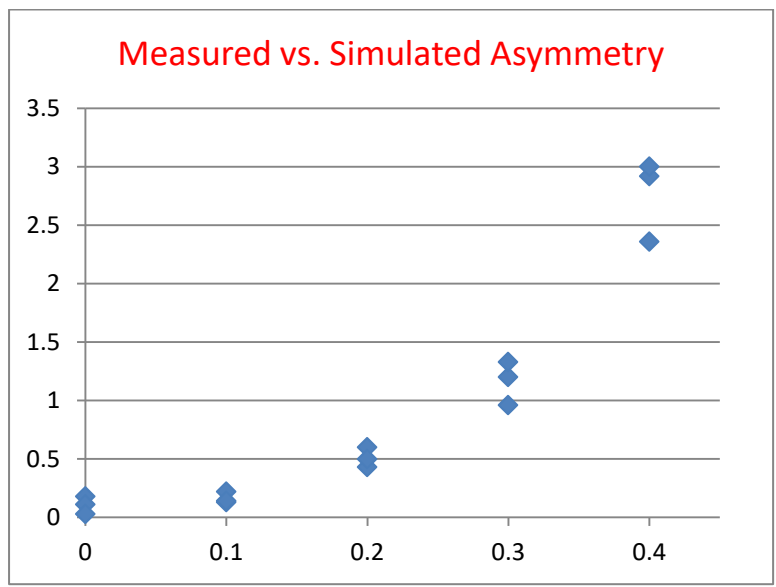

**Figure 9:** Comparison of *simulated* asymmetry (horizontal axis) and *measured* asymmetry (vertical axis) in videos of cell-reproduction processes (mitotic spindles).

*5.3. Different levels of statistics and misinterpretation of "standard deviation"*

The negative connotation of the term *deviation* and the consequent presumption of "standard deviation should be small" can lead to problematic quantification methodologies and their interpretations in tracking problems.

The statistical distribution of some data points is not necessarily Gaussian-looking. It can be asymmetric, sparse … As such, a correct measurement may have a *variance* much larger than (the square of) the mean value. A large variance or standard "deviation" can only be problematic if originating from measurement errors of a quantity known to be constant. In other cases, and especially in the case of tracking microtubule-associated end-binding proteins, it is important to clarify what "standard deviation" means, before dealing with its numerical value.

Figure. 10 shows three choices of data points (*statistical ensembles*), which can be used in calculations. In the left panel ("Track-Level" statistics), the output is an array of mean values (and standard deviations). In the middle panel ("Image-Level" statistics), the output is also an array (with fewer elements). In the right panel ("Group-Level" statistics), the output is just a number.

Further averaging of the arrays produced in the left and the middle panels can generate almost the same average value, as generated in the right panel. However, the standard deviation is expected to assume (and based on real-life images does assume) different values, depending on which type of statistics and how many averaging are employed.

In comparative studies (of say, normal and mutated cells), Track-level statistics may show (nearly-) similar standard deviations at Group Level, but considerable differences at Track Level. The temptation of dealing with small standard deviations simply filters out this important piece of information.

In the case of particle velocity, such large variances affect the estimated "displacement" and hence the estimated displacement-to-distance-ratio and the validity of tracking in the first place.

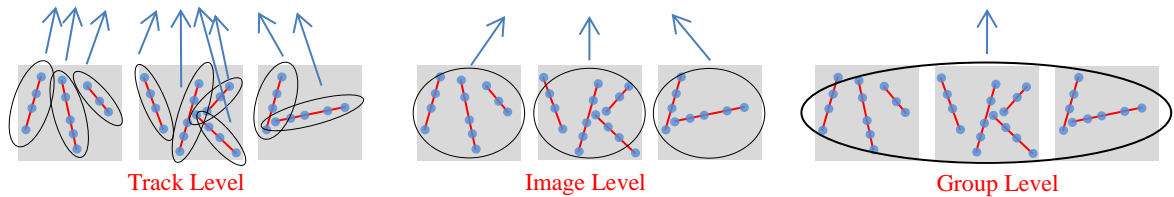

**Figure. 10:** The histogram of intra-pair distances (and the indices defined with it) can be calculated at Track Level, Image Level, or Group Level.

## 5.4. Pairing and pair analysis

Pair formation is based on identification of the closest neighbor of a given spot (in the same frame) and limiting this *intra-pair distance* to a threshold. Each unpaired spot is considered as a (degenerate) pair of overlapping spots.

### 5.4.1. Track - (in)dependent parameters

Properties such as the rate of switching between "Small" and "Large" distances are related to specific cells. They can be calculated only with explicit knowledge of (consecutive pairs along) different tracks. These properties are affected by uncertainties in tracking (spurious merge or split of a cell track, handling mitotic events ...). Other properties (such as the mean intra-pair distance, or the fraction of pairs with intra-pair distances above a threshold) can be defined not only at *track level*, but also at *image level* and *group level*, as depicted in Figure. 10.

### 5.4.2. Different (statistical) levels of analysis

By using all pairs in an image, one can form an *image histogram* (of intra-pair distances) and define indices such as the mean (intra-pair) distance, or the fraction of pairs with (intra-pair) distance above a threshold. By performing this analysis on different images in different groups, one can get a set of indices and hence some scatter plots. Each point in such a scatter plot represents a time-series image.

The aforementioned procedure may also be done by combining all pairs of all images within a group to form a *group histogram*. Scatter plots in this case will have only one point per group.

### 5.4.3. Static properties

A typical histogram of (intra-pair) distances has a large value (almost 50% of all counts in real-life images) at "zero". Here, a distance of zero is larger than the resolution distance of optical imaging and depends on the distinguishability of detected spots. Such histograms are neither intuitive to visualize, nor simple to analyze directly. However, if we integrate (add up consecutive values of) the histogram function, we get an exponential-looking increasing curve (Mathematically speaking, integration of the *Probability Distribution Function*; PDF generates the *Cumulative Distribution Function*; CDF). By normalizing this integrated histogram to the total number of counts, we get a comparable basis (irrespective of the number of data points).

Two main indices that can be calculated using the integrated histogram are 1) the fraction of intra-pair distances greater than a threshold (1 - "y-intercept"), and 2) the *characteristic distance*; estimated by fitting an exponential function to the curve (at distance values greater than the threshold). Preliminary automated analysis of different types of cells has shown some noticeable changes in these parameters across groups of cells.

Investigation of real-life images show the relevance of the exponential model and a good fit in most cases for image-level and group-level statistics. In track-level statistics, the histogram corresponds to too few points to be meaningful for model-based fitting. A simpler index like the median value is more meaningful in this case.

*5.4.4. Dynamic properties*

The dynamics of intra-pair distances *along a specific track* can be quantified by counting the number of times the distance exceeds or falls below the threshold distance. By dividing this number to the number of frames, this "effective rate" becomes comparable for tracks with different lengths. The inverse of this effective rate is the *effective period*. (This definition does not differentiate between transitions from "Small" to "Large" and from "Large" to "Small". Otherwise, the effective period would be larger with a factor of two).

In image-level and group-level statistics, there will be many such indices corresponding to different tracks. Thresholds for the number of points along a track and also an outlier-robust index (median) for the statistics of the obtained values have been considered.

*6. Concluding remarks*

Problem-specific meanings and requirements of "tracking" have been identified and addressed in select cell-biological problems (centrosomes, microtubule-assisted end-binding proteins, and diffusive biomolecules). The resulting solutions are available as free and open-source Java programs; to be used with the common free bio-image analysis software *ImageJ*. Intuitive interpretation and validation of results by the end-user are associated with extensive behind-the-scene mathematical and programming details. These details have been publicly documented for further assessment and development, without distracting the end-user.

*Appendix I: End-User's Guide*

*AI.1 Pre-processing*

Multiple 2D and 3D filters (part of the ImageJ/Fiji API) are offered to the user, as shown in Figure 11. The user may first choose one, some, or all of the filters. Then, the parameters of the filter can be set interactively by observing the result of applying multiple filters with the selected sequence and selected parameters. Any selected filter can also be deactivated by selecting its (first) parameter to the minimum value.

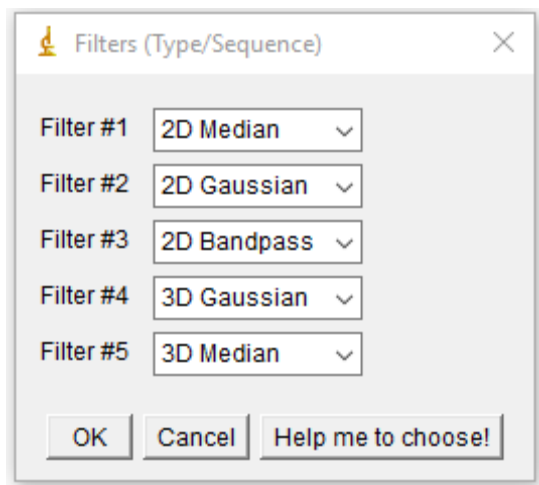

**Figure 11:** Type and sequence of filters to be used interactively in the pre-processing step.

*AI.2 Particle Analysis*

Particle Analysis includes two different modes, as shown in Figure 12:
- In the *Quantitative* (slider) mode, parameters such as size, circularity, and intensity of the sought particles can be adjusted interactively, and the detection results can be seen in real-time. The so-obtained optimal parameters can be then used for automated tracking.
- The *Object-based* mode offers the possibility of interactive selection of ROIs with mouse-clicks. At the core of the Java code is an ImageJ Macro that performs the same operation. The Java code will incorporate other similar Macros in the upcoming versions.

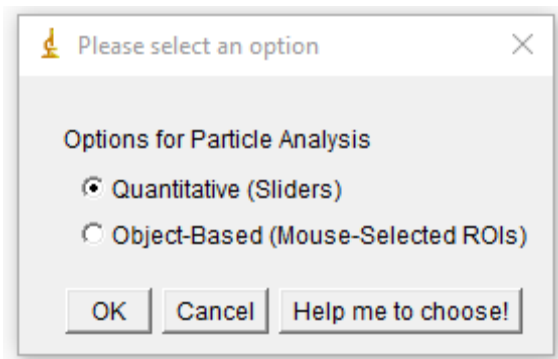

**Figure 12:** The two modes of Particle Analysis

*AI.3 Pairing, Tracking, and Group Analysis*

- *Pairing:* In cases such as centrosome pairs, the images should be first analyzed to identify and to associate pairs of particles, as shown in Figure 3. Conventional tracking can be done only afterwards on detected pairs (not original individual particles). It is also necessary to keep track of the association between original particles and linked pairs for later quantification of static and dynamic pair properties (mean intra-pair distance and the "near-far" switching rate, for instance).

- *Tracking:* Tracking can be done using two methods; Fiji's own powerful *TrackMate* algorithm, or an algorithm developed here and referred to as "Real-Time". The latter is simply based on linking the closest particles *across adjacent frames*, and is also used for pairing objects *in the same frame*. In Multi-Channel Kinetics, a cell (or a mitotic spindle) is followed in one channel and a measurement over a *correspondingly-positioned ROI* is made on a second channel.
- *Group Analysis:* Using this option, one can apply an automated tracking and track analysis scheme to multiple images and save all the results with pertinent names. Furthermore in performing Asymmetry or Pair analysis, all these results can be squeezed in a clear 2D scatter plot with each point representing two indices (associated with a time-series image) and the images from the same group being represented with the same color.

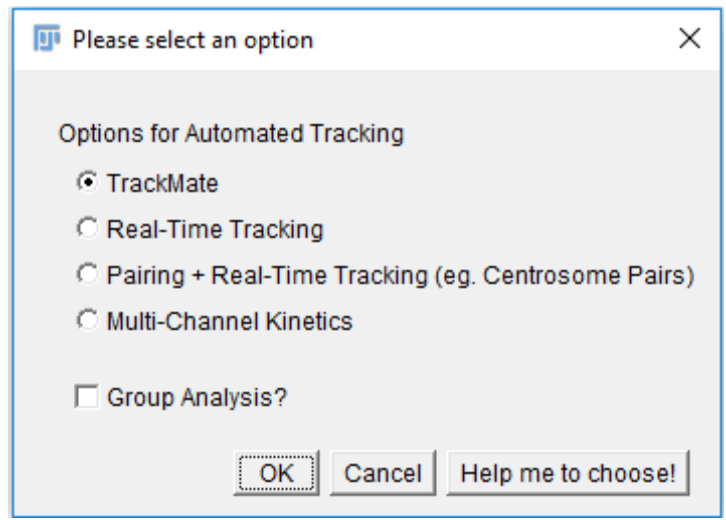

**Figure 13:** Different options for Pairing, Tracking, and Group Analysis

Once a tracking method is chosen, method-specific parameters of tracking should be adjusted. Here it is also possible to choose Asymmetry and/or Diffusion Analysis (Ruthardt, 2011) to analyze the tracks. Figure 14 shows the parameters that can be adjusted prior to tracking with *TrackMate*. In the case of Pair-tracking, the (average) static and dynamic properties of pairs will be automatically calculated and displayed.

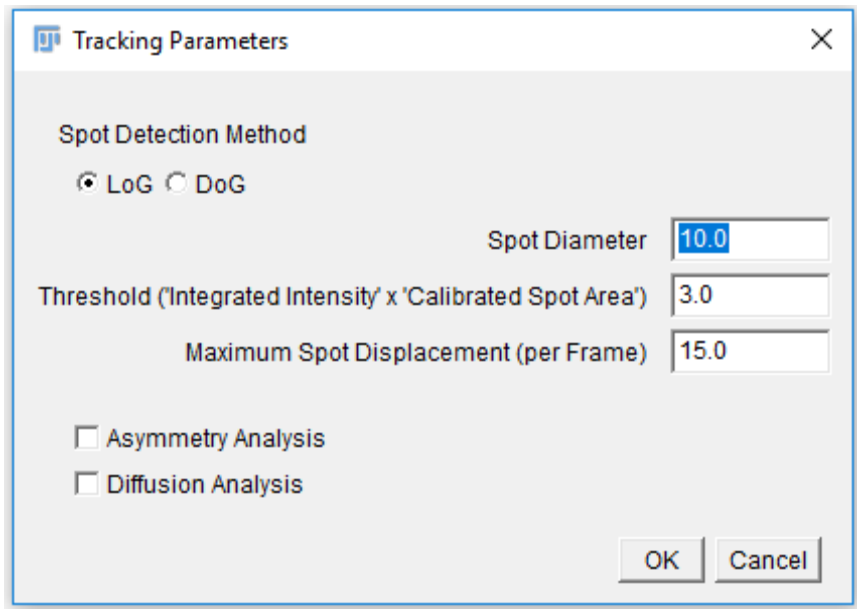

**Figure 14:** Important Parameters affecting the performance of *TrackMate* should be adjusted by the user.

*AI.4 Interactive track visualization and analysis*

There are four major interactive visualization schemes (with associated analyses), as shown in Figure 15:

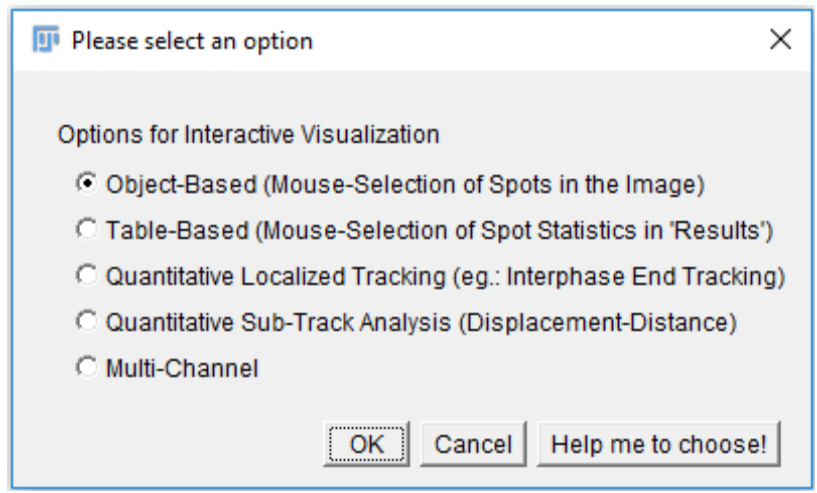

**Figure 15:** Four modes of interactive visualization (and associated analyses)

- In *object-based* mode, by clicking on an arbitrary point in a time-series image, the closets particle to that point will be highlighted. The particle will also be highlighted in all other frames (all spots of the corresponding track are highlighted). The color-coding of high-lighting is based on the track index (the first track in dark blue and the last one in dark red).
- In the *table-based* mode, the spots will be displayed (and followed) while browsing tracking results in the "Results" table.
- In *quantitative localized tracking*, one tries to select a subset of tracks that emerge from the same source. It makes it possible to achieve single-cell statistics out of tracks generated by multiple cells.

- *Quantitative sub-track analysis* allows the user to split the tracks into virtual sub-tracks and to analyze the "displacement-distance" of all subtracks. Such an analysis differentiates between "fast" and "slow" particles and also between "linear" and "zigzag" paths. All parameters, including the critical length of sub-tracks can be adjusted interactively and the associated analysis results will be seen in real-time.
- In Multi-Channel visualization, one channel of an RGB image is fixed, and the other channel can be translated, rotated, and (anisotropically) scaled.

*AI.5 Simulation*

Different options for simulation of an image (or z-stack or time-series) are shown in Figure 16:

- Realistic simulation of particles with Gaussian-looking densities, with arbitrary lengths of major axes, and arbitrary orientations (in 2D or 3D)
- Simulation of the (static) mitotic spindle (in 2D or 3D)
- Time-series simulation of the dynamic cell-reproduction process (2D Spindle) or the stage preceding it (*Interphase*)
- Time-series simulation of moving spindles in one channel and associated spots with time-varying intensities in another channel (Two-Channel Spindle)
- Time-series simulation of (centrosome) pairs with adjustable statistical properties of pairs (dialog-box) and adjustable pair trajectories (mouse-clicks)
- Simulation of sinusoidal variations of intensity with gradual increase of the rate of oscillations. Concepts such as resolution, deconvolution, and (imperfect) super-resolution can be intuitively and interactively seen.

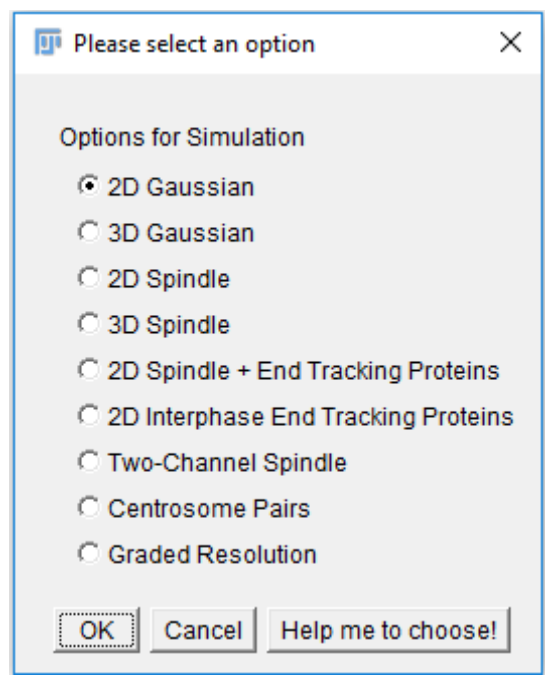

**Figure 16:** Different options for simulation of an image (including z-stacks or time-series)

*Appendix II: Practical issues in asymmetry analysis*

*AII.1. Resolution of the orientation histogram (bin size)*

Finding the dominant orientation based on the maximum of the histogram is unreliable, as the peak value depends critically on the bin size. More importantly, for small bins, the baseline of the orientation histogram can be zero, and it makes the orientation index infinity.

Angular bins also rely implicitly on angular resolution in images (on a polar grid, and not on the conventional Cartesian grid of camera pixels). The angular resolution is non-uniform, and less reliable for small lengths. As such, increasing the number of bins can introduce angular aliasing (spurious oscillations in orientation histogram) at very large values of the bin number.

With measured images, we have found a bin number of ~20 good enough for fast generation of smooth orientation histograms (Figure 2) with distinguishable peaks. A bin number of 1000 shows spurious-looking oscillations (indicative of aliasing). In case more angular resolution is required, we would suggest a maximum bin number of ~100.

In analysis of a single image, the number of bins (when large) may critically affect the orientation index. However, in analysis of a series of images with the same (small) number of bins, this issue is not critical.

*AII.2. Histogram at boundaries (close to ±180 °)*

There are no angles greater than 180 ° or smaller than -180 °. As such, the histogram bins *centered* at the boundary values of -180 and 180 degrees are only meaningful over half of their angular span. Furthermore, due to circular symmetry, the two angles -180 ° and 180 ° are indeed the same. As such, we add the histogram values corresponding to these two boundary angles and assign it to both. It forces circular symmetry and is intuitive for visualization. The redundant data point is however excluded from histogram analyses (dominant orientation identification).

*AII.3. Circular shift for identification of the dominant direction*

While searching for the dominant orientation, the area under the histogram around all angles (histogram bin centers) should be calculated. This area should be calculated over a width of 180° and compared with the rest of the histogram. When the required angular range exceeds *±180°*, a value of *360°* is subtracted (or added). This *phase-wrapping* enforces circular symmetry to bring the angles back to the available histogram range of $[-180, 180)$.

*AII.4. Manual optimization of tracking parameters*

The program *TrackMate* offers many more parameters for general-purpose tracking problems. It might be seen as flexibility or uncertainty, depending on circumstances. In our case (reliable identification of dense tracks with high frame-rate videos), some parameters have been eliminated by forbidding gaps or split/merge events. Many parameters for filters on tracks have also been eliminated, as one can enforce them later by post-processing *TrackMate* results (and not necessarily *by TrackMate* itself).

*AII.5. Robustness to parameters of the tracking algorithm*

Inappropriate parameters of the tracking program may result in spurious merge of two adjacent tracks, or a spurious split of a single track. This can affect parameters such as duration or length of the track. However, calculations based on links (such as those for *Directionality* and *Asymmetry* indices) are less vulnerable to such spurious merge/split events.

Our high frame-rate imaging enhances this robustness even further. With ~900 tracks and ~36,000 links per image, there is an average of ~40 links per track. In the case of a spurious split (or merge), only one data point will be removed from (or added to) the 40-element array.

*AII.6. Insensitivity of Directionality and Asymmetry indices*

Defining figures of merit as ratios (Figure 2) makes them nearly-independent of the *exact* numbers of tracks and frames, as A0, A1, and A2 will be all scaled in nearly-similar ways. The numbers of tracks and frames; however, should be large enough to be representative ensembles of the sought statistics.

*AII.7. Deviations from a linear track shape*

Identification of individual spots is based on the assumption of a circularly-symmetric local intensity distribution for *centering*. Morphological changes of the spot intensity profile from one frame to the next can affect the accuracy of centering (a skewed profile). But can deviations from a linear track be only attributed to such centering errors? A more conclusive answer comes from indirect visualization of tracks with time-projections, in which case no explicit spot detection is used. Investigation of track shapes in such cases also suggests some level of deviation from a linear track shape. So, this effect is (at least to some extent) authentic and not due to numerical errors of tracking. Disentangling possible factors such as 2D nonlinear movements, out-of-plane (3D) movements, or microscope vibrations requires further investigation.

*Appendix III: Mathematical/Algorithmic details*

*AIII.1. Linking (Related classes: "Tracking" and "Linking")*

At the core of all three problems of single-particle tracking, pairing, and pair-tracking is the problem of linking two images. In tracking, these two images are different, and the closest point in one image to a given point in the other image is sought. In pairing, the two images are the same, and the closest non-trivial point (i.e.; the second closest point) is sought. In pair-tracking, linking is employed twice; once to pair the spots in a given frame and once to link pairs across frames.

Identification of closest neighbors is based on the calculation of all distances and sorting them. Let the two images be denoted by $I_1$ and $I_2$; *unsorted* spots in an image $I$ by $\mathbf{S}_I = \{S_I^1, S_I^2, S_I^3, \ldots\}$; the number of spots in an image $I$ by $n(\mathbf{S}_I)$; and the distance matrix of two images $I_1$ and $I_2$ as $D(I_1, I_2)$; and the distance between two spots $S_1$ and $S_2$ as $\|S_1 S_2\|$. Then we have

$$\mathbf{D} = \mathbf{D}_{n(\mathbf{S}_{I_1}), n(\mathbf{S}_{I_2})} = \{D_{i,j}\} = \{\|S_{I_1}^i S_{I_2}^j\|\}$$

In tracking, a link from a spot $S_{I_1}^i$ to a spot $S_{I_2}^j$ is valid, if only

- $S_{I_2}^j$ is the closest point (in $I_2$) to $S_{I_1}^i$;
- $S_{I_2}^j$ has not been already assigned to another point $S_{I_1}^k$; and
- $\|S_{I_1}^i S_{I_2}^j\|$ does not exceed the user-defined *threshold for displacement*.

Linking objects between the two images starts from the smallest $D_{i,j}$. At the end, any unlinked spot $S_{I_2}^j$ is considered as the first point of a new track.

In pairing, a pair formed with $\{S_I^i, S_I^j\}$ is valid, if only

- $S_I^j$ is the closest point (in $I$) to $S_I^i$;
- $S_I^j$ has not been already assigned to another point $S_I^k$; and
- $\|S_I^i S_I^j\|$ does not exceed the user-defined *threshold for intra-pair distance*.

Pairing objects starts from the smallest non-zero $D_{i,j}$. At the end, any unpaired spot $S_I^j$ is considered as a degenerate pair (two overlapping spots).

*AIII.2. Quantitative localized tracking (Related classes: "VisualizationTracksInterphaseEB")*

In localized tracking (Figure 5), two distance parameters and also a "maximum divergence angle" are employed to attribute a track to one specific source. This angle is defined as shown in Figure 17.

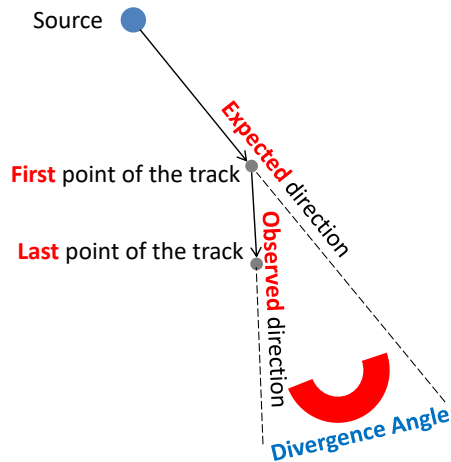

**Figure 17:** Calculation of track divergence (for source association) based on the locations of the source and the first/last points of the track.

*AIII.3. Simulation of a (3D) Gaussian spot (Related classes: "VisualizationTracksInterphaseEB")*

Let the image spatial coordinate be denoted by $r = (x, y, z)$; the image intensity at each point $r$ by $I(r)$; the 3D rotation matrix corresponding to an angle of $\theta$ around a principal axis $a$ ($a = x$ or $y$ or $z$) by $R_a(\theta)$; and a (unity-amplitude) Gaussian function with a width parameter $\sigma$ by $g_\sigma(p) = e^{-p^2/2\sigma^2}$.

With the user-defined position vector $r_0 = (x_0, y_0, z_0)$; 3D rotation matrix $R$ parameterized with ZYZ euler angles $(\alpha, \beta, \gamma)$; anisotropic widths $(\sigma_x, \sigma_y, \sigma_z)$; and maximum intensity $I_{max}$, the simulated intensity of the z-stack at a point $r$ is given by $I(r) = I_{max} \, g_{3D}(R'(r - r_0))$, where $g_{3D}(x, y, z) = g_{\sigma_x}(x) g_{\sigma_y}(y) g_{\sigma_z}(z)$ and $R = R_z(\gamma) R_y(\beta) R_z(\alpha)$. The 2D Gaussian function is synthesized in a similar and simpler form with only 2 width parameters (rather than 3), two center position parameter (rather than 3), and 1 rotation parameter (rather than 3).

*AIII.4. Simulation of (3D) cell-reproduction trajectories (Related classes: "MTTrackSet3D")*

Let the image spatial coordinate be denoted by $r = (x, y, z)$; the two end-points (particle sources) by $P_1$ and $P_2$; the mid-point of the <u>3D arc</u> joining $P_1$ to $P_2$ by $H$; the mid-point of the <u>line</u> joining $P_1$ to $P_2$ by $M$; the unit-vector along $P_1P_2$ by $\hat{n}$; the length of $MH$ by $r$; and the origin of the coordinate system by $O$ ($x = 0, y = 0, z = 0$). One can write:

$$\overrightarrow{OM} = (\overrightarrow{OP_1} + \overrightarrow{OP_2})/2$$
$$\hat{n} = (\overrightarrow{OP_2} - \overrightarrow{OP_1})/\|\overrightarrow{OP_2} - \overrightarrow{OP_1}\|$$

Different points $H$ of different tracks lie on a circle (with a radius $r$) perpendicular to $P_1P_2$ in the middle (at $M$). By parameterizing this circle with an *independent* angle $\psi$ (and using a *dependent* angular parameter $\theta$), one can write:

$$\overrightarrow{OH} = \overrightarrow{OM} + \overrightarrow{MH} = \overrightarrow{OM} + r(\sin(\theta)\cos(\psi), \sin(\theta)\sin(\psi), \cos(\theta))$$
$$\tan(\theta) = -(\hat{n}.\hat{z})/[(\hat{n}.\hat{x})\cos(\psi) + (\hat{n}.\hat{y})\sin(\psi)]$$

Different tracks are generated by scanning the parameter $\psi$ over a range of $2\pi$ (and finding the corresponding parameter $\theta$). In the implemented program, three sliders let the user limit the angular span of $2\pi$ to a smaller range, and also choose the number of $\{\psi_n\}$ points (number of tracks).

Once the three points $\{P_1, H, P_2\}$ of a track (microtubule) are known, intermediate points can be found using second-order interpolation; i.e., a quadratic Bézier curve. Describing the curve with a parameter $t$ so that $P(t = 0) = P_1$, $P(t = 0.5) = H$, and $P(t = 1) = P_2$, one can write:

$$P(t) = (1-t)^2 P_1 + 2t(1-t)H + t^2 P_2$$

A similar procedure is used for the 2D case (the class *"MTTrackSet"*), and its dynamic version (the class "*Simulation2DSpindleEBProteinsCore*")

*AIII.5. Simulation of videos of the cell-reproduction process (Related classes: "EBs")*

Generation of tracks is similar to the static case of frming a 2D spindle. In the dynamic case, there are trivial new functionalities for creating a timer and responding to timer events. The major nontrivial new functionalities here have been implemented in the class *EBs* for generation of particles at appropriate frames (with specific densities and rates); and also for displacement and removal of particles in subsequnt frames.

A full track (microtubule) can be modelled with 1) three points $\{P_1, H, P_2\}$ and 2) a scalar parameter $0 \leq t \leq 1$. Moving particles on a track can be modelled purely with the scalar parameter $t$. The actual synthesis (rendering) of particles will be done as a separate task later by combining the information of 1) absolute track coordinates $\{P_1, H, P_2\}$ and 2) parametric coordinates of particles.

Four parameters are used to model a particle: the $t$ parameters of the first and the last points ($t_1$ and $t_2$); the direction of the particle (whether it emanates from $P_1$ or $P_2$); and the duration of the particle (number of frames, over which the particle is visible).

For particles emanating from $P_1$, the range of the curve parameters is $0 \leq t_1 \leq t_2 \leq 0.5$. These two parameters increase from one frame to the next until the particle vanishes. Such a particle vanishes, when updating $t_2$ results in a value larger than the limit (0.5), or when the maximum duration of track is reached. Similarly, for particles emanating from $P_2$, the range of parameters is $0.5 \leq t_1 \leq t_2 \leq 1$. These two parameters decrease from one frame to the next until the particle vanishes. Such a particle vanishes, when updating $t_1$ results in a value smaller than the limit (0.5), or when the maximum duration of track is reached. In the presence of *asymmetry*, the limit value is added with a small random number, the magnitude of which is determined by the level of asymmetry (set with a slider).

Vanishing a particle is implemented by setting its $t$ parameters to invalid (out of range) values. Upon updates, such *outgoing* particles are replaced with new *incoming* particles if there is room between the source ($P_1$ or $P_2$) and the current closest particle to the source (to avoid overlap or "collision" of particles). In rendering, only particles with valid values of $t_1$ and $t_2$ will be displayed.

*AIII.6. Simulation of double-spots (Related classes: "SimulationCentrosomePairs" and "PairCenters")*
The points specified by the user (mouse-clicks) are interpolated linearly inbetween. The (interpolated) points specify centers of pairs. For a given center of pair, the actual visible spots are generated based on the user-specified maximum intra-pair distance and randomly-generated scale factors.

Let the image spatial coordinate be denoted by $\boldsymbol{r} = (x,y)$; the user-defined points of a trajectory by $\boldsymbol{P} = \{p_1, p_2, p_3, \ldots, p_N\}$; and the corresponding frame indices of these points by $\boldsymbol{F} = \{f_1, f_2, f_3, \ldots, f_N\}$. Also let the image intensity at each point $\boldsymbol{r}$ at frame $f$ be denoted by $I_{pairs}(\boldsymbol{r}, f)$; and the intensity distribution function of a single spot by $g(\boldsymbol{r})$. In its simplest form, $f(\boldsymbol{r})$ corresponds to a single point (in a continuous formulation or a single pixel in a practical discrete formulation); i.e., $g(\boldsymbol{r}) = I_0 \delta(\boldsymbol{r})$. For simplicity and clarification of equations, we start with the simpler situation of single spots at pair centers, instead of pairs. For this auxiallry situation, the image intensity function $I_{centers}(\boldsymbol{r}, f)$ can be written as the contribution of all user-defined spots and the interpolated points inbetween:

$$I_{centers}(\boldsymbol{r},f) = g(\boldsymbol{r} - \boldsymbol{r}_{p_N})\delta_{f,N} + \sum_{i=1}^{N-1} \sum_{j=f_i}^{f_{i+1}} g(\boldsymbol{r} - [t_{i,j}\boldsymbol{r}_{p_i} + (1-t_{i,j})\boldsymbol{r}_{p_{i+1}}])\delta_{f,j}$$

Here $\delta_{f,j}$ is Kronecker's delta function (1 for similar indices and 0 otherwise); and $t_{i,j} = (j - f_i)/(f_{i+1} - f_i - 1)$ is the interpolation argument of the linear Bézier curve connecting the points $p_i$ and $p_{i+1}$ together.

In transition to the actual case of pairs, each term in $I_{centers}(\boldsymbol{r}, f)$ is replaced with two; and the centers are slightly and randomy displaced from the center coordinate. Assuming a maximum intra-pair distance of $d$ and also four random variables with given distributions $\boldsymbol{S} = \{n_1(s), n_2(s), n_3(s), n_4(s)\}$, the sought intensity distribution function of pairs can be written as:

$$I_{pairs}(\boldsymbol{r},f) = \sum_{k=1}^{2} \left[ g(\boldsymbol{r} - \boldsymbol{r}_{S,k,p_N})\delta_{f,N} + \sum_{i=1}^{N-1} \sum_{j=f_i}^{f_{i+1}} g(\boldsymbol{r} - [t_{i,j}\boldsymbol{r}_{S,k,p_i} + (1-t_{i,j})\boldsymbol{r}_{S,k,p_{i+1}}])\delta_{f,j} \right]$$

Here, a randomized vector is $\boldsymbol{r}_{S,k,p} = \boldsymbol{r}_P + (d/2)(n_1(s), n_2(s))\delta_{k,1} + (d/2)(n_3(s), n_4(s))\delta_{k,2}$.

Note that the scalar argument $s$ in $n_1(s)$ merely denotes the randomness of the variable $n_1$. In practice, it my be attributed to the index of the pseudo-random number generated by the program.

Enforcing a hard limit on intra-piar distances is possible when employing a distribution with a limited range (such as the uniform distribution). In using a distribution with a (theoretical) unlimited range (such as the normal distribution), one should clip the range manually. This clipping should be considered while interpreting statistical properties of the original distribution.

Changing the distribution of random numbers is an easy programming task; yet provides invaluable insight into the "randomness" of such processes at a more quantitative level.

*AIII.7. Simulation of graded resolutions (Related classes: "Simulation2DGradedResolution")*
Let the image spatial coordinate be denoted by $(x, y)$; the width of the image by $w$; the chirp-factor by $c$; and the intensities at the leftmost, midle, and rightmost pixels by $I_1$, $I_2$, and $I_3$. The intensity $I(x, y)$ at an arbitrary point is calculated as follows:

$$x_1 = 0, x_2 = w/2, x_3 = w$$
$$\alpha = [(I_1 - I_2)(x_3 - x_2)^2 - (I_3 - I_2)(x_1 - x_2)^2]/[(x_1 - x_2)(x_3 - x_2)(x_3 - x_1)]$$
$$\beta = [(I_1 - I_2) - \alpha(x_1 - x_2)]/[(x_1 - x_2)^2]$$
$$I(x, y) = [I_2 + \alpha(x - x_2) + \beta(x - x_2)^2] |\cos(cx^2)|$$

*Appendix IV: Architectural considerations*

The architecture used for modelling and identification of tracks was a simple one, including classes for *Spot*, *Pair*, *Track*, *TrackSet*, *SpotDetection*, *Linking*, and *Tracking*, as shown in Figure 18. The *Mat* and *SimpleBlobDetector* classes correspond to the C++ API OpenCV, and the corresponding classes in ImageJ API are *ImagePlus* and *ParticleAnalyzer.*

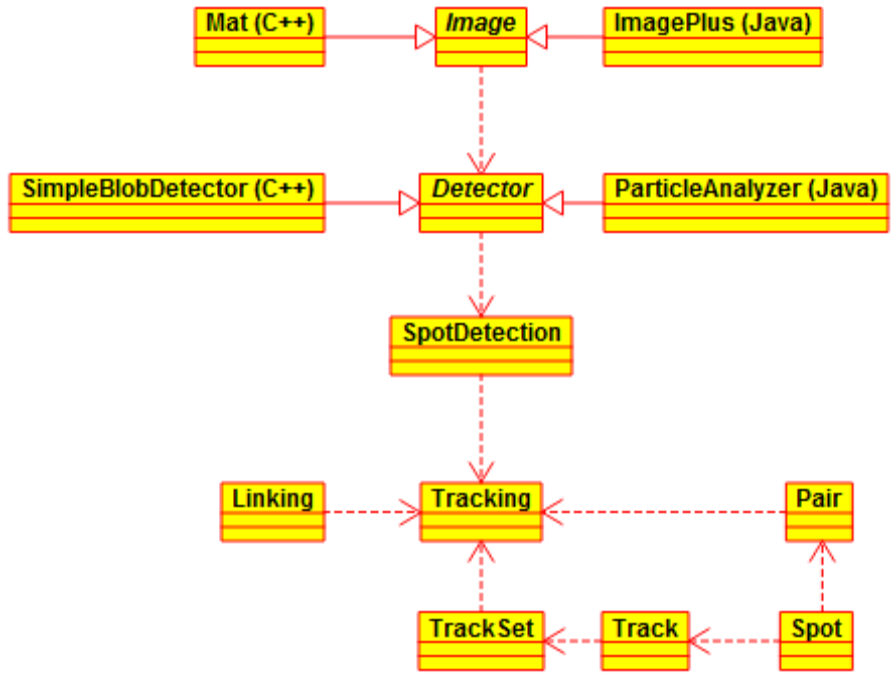

**Figure 18:** Simple architecture of the classes used for modeling and identification of tracks

In the course of modifying the sample program *Cross_Fader.java* (Schmid, 2014) to develop interactive visualization/simulation programs, some patterns had to be repeated (saving image properties, handling sliders for hyperstacks …). Also by increasing the number and the type of GUI elements, handling GUI data (minimum values, maximum values, initial values, and the labels of all sliders, for instance) required dedicated functionalities (classes). The original and some modified architectures have been shown in Figure 19 and Figure 20, respectively.

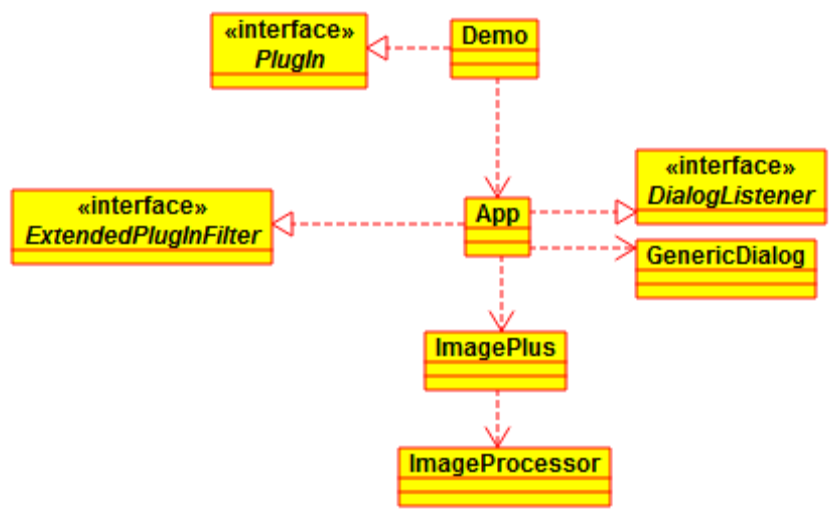

**Figure 19:** Original architecture used for interactive visualization/simulation

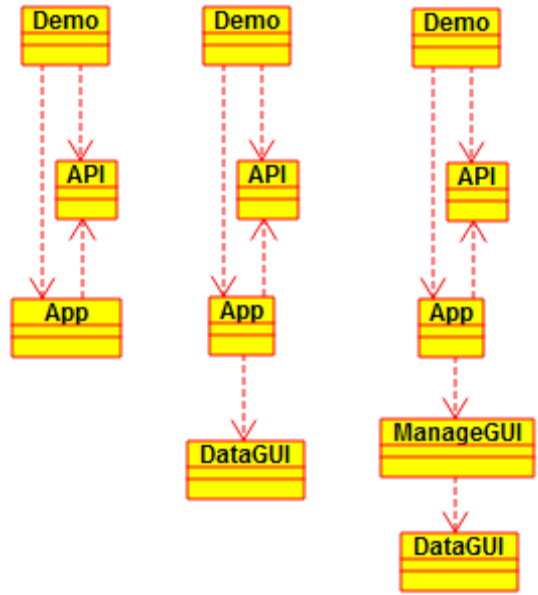

**Figure 20:** Modified architectures used for interactive visualization/simulation: (Left) Simplified architecture shown in Figure 19; (Middle) Incorporating all GUI data (type, label, minimum value, maximum value, current value … of all sliders, radio buttons …) in a new dedicated class *DataGUI*; and (Right) Incorporating all "callback functions" in a new dedicated class *ManageGUI*. The goal is to increase the modularity, simplicity, and reusability of the *App* (and associated) class(es).

A more detailed architecture of a typical application is exemplified in Figure 21 for the case of tracking *Multi-Channel Kinetics*. The main functionality is managed by the class *FrapAnalysis*, and two interactive analyses are realized with the classes *Analyze* and *ParticleAnalysisQuantitative* (and their associated classes).

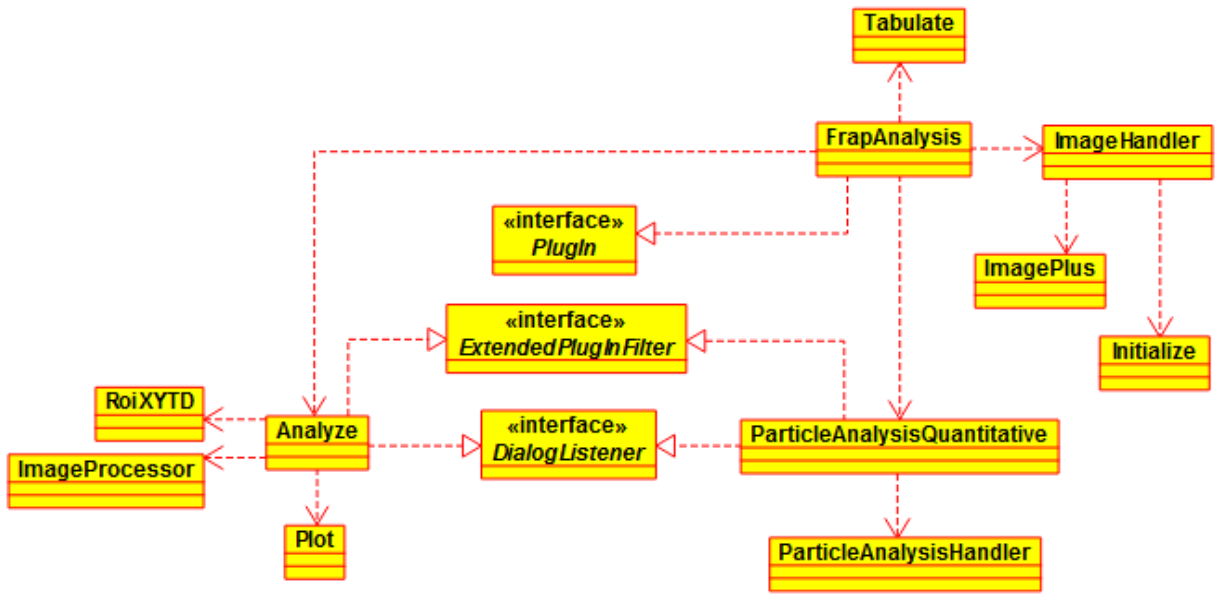

**Figure 21**: Multi-Channel Tracking managed by the class FrapAnalysis and using two classes Analyze and ParticleAnalysisQuantitative for two channel-dependent interactive adjustment of parameters.

23